\documentclass[twocolumn,prl,preprintnumbers,superscriptaddress,amsmath,amssymb]{revtex4}      
\usepackage{amsmath,dcolumn,graphicx,epsf,ulem}       
\setkeys{Gin}{width=8.5cm,keepaspectratio}       
\usepackage{dcolumn}% Align table columns on decimal point       
\usepackage{bm}% bold math       

\newcommand{\ee}[1]{\times 10^{#1}}    
       
\begin{document}       
       
\title{Magnetism in nanoparticle $\rm LaCoO_3$}       
      
\author{A.~M.~Durand}      
\affiliation{Department of Physics, University of California, Santa Cruz, CA 95064, USA}      
\author{D.~P.~Belanger}      
\affiliation{Department of Physics, University of California, Santa Cruz, CA 95064, USA}      
\author{F.~Ye}      
\affiliation{Quantum Condensed Matter Division, Oak Ridge National Laboratory, Oak Ridge, Tennessee 37831, USA}      
\author{S.~Chi}      
\affiliation{Quantum Condensed Matter Division, Oak Ridge National Laboratory, Oak Ridge, Tennessee 37831, USA}      
\author{J.~A.~Fernandez-Baca}      
\affiliation{Quantum Condensed Matter Division, Oak Ridge National Laboratory, Oak Ridge, Tennessee 37831, USA}      
\author{C.~H.~Booth}         
\affiliation{Chemical Sciences Division, Lawrence Berkeley National Laboratory, Berkeley, CA 94720, USA}   
\author{M.~Bhat}      
\affiliation{Castilleja School, Palo Alto, CA 94301, USA}

\date{\today}      
   
%%% Corrections put in by Alice on 9/28/13   
% For the subscripts, let's do T_C (capital C)   
% and gamma_c (lowercase c)   
% and T_o (lowercase o)  
       
\begin{abstract}       
%Rewrote the abstract      
   
$\rm LaCoO_3$ (LCO) nanoparticles were synthesized and their    
magnetic and structural properties were examined using    
SQUID magnetometery and neutron diffraction. The nanoparticles    
exhibit ferromagnetic long-range order beginning at $T_C \approx 87$~K   
that persists to low temperatures. This behavior is contrasted    
with the ferromagnetism of bulk LCO, which also starts at $T_C \approx 87$~K   
but is suppressed below a second transition at $T_o \approx 37$~K. %due to a structural phase transition.   
The ferromagnetism in both systems is attributed to    
the tensile stress from particle surfaces and    
impurity phase interfaces. This stress locally    
increases the Co-O-Co bond angle $\gamma$. %and competes with the thermal contraction of the lattice.    
It has recently been shown that   
LCO loses long-range ferromagnetic order    
when $\gamma$ decreases below the critical value    
$\gamma_c=162.8$$^\circ$. Consistent with this    
model, we show that $\gamma$ in nanoparticles remains    
larger than $\gamma_c$ at low temperatures, likely a    
consequence of all spins being in close    
proximity to surfaces or interfaces.     
       
\end{abstract}        
       
\maketitle       
       
The magnetism of $\rm LaCoO_3$ (LCO) is well-known to    
be unusual.  Both ferromagnetic and antiferromagnetic interactions   
have been proposed in the bulk material.~\cite{g58,akg01} %%%  
The dominant interaction above $T=100$~K is antiferromagnetic; however,  %%% 
in our previous study, we did not find evidence of  
antiferromagnetic ordering upon decreasing $T$.~\cite{dbbycfb13} %%% 
We thus concluded %%% 
that the antiferromagnetic interactions are strongly frustrated. %%%   
We have also recently shown~\cite{dbbycfb13} that    
weak ferromagnetic order occurs below $T_C=87$~K in small applied fields, $H \le 100$~Oe,   
yet is lost below $T_o \approx 37$~K. %%% we'll stick to 37 rather than 40    
Studies of LCO nanoparticles~\cite{fmmpwtvhvg08,zhzgzs09,zszhyz07,wzwf12}   
and thin films~\cite{fpssnssmrl07,hrsd09,fapssl08,srkkmsw12,fdaeaeksgl09,fms08,prkkmsfw09,pfmwalsns08,pbsskzwld11,rhsd10} also   
show the presence of a ferromagnetic phase transition near $87$~K, though    
these materials do not exhibit the same loss of ferromagnetic long-range   
order seen in bulk LCO below $37$~K.   
   
The unusual low temperature magnetism in the bulk    
has been shown to correlate well with the behavior    
of the Co-O-Co bond angle, $\gamma$.   
Experimental measurements~\cite{dbbycfb13} indicate that   
ferromagnetism only exists when $\gamma$ is   
greater than a critical value $\gamma_c= 162.8^\circ$, which happens only for $T>T_o$. %%%  
Above $T_o$, all of the lattice parameters, including $\gamma$, show   
power-law behavior in $T-T_o$.  For $T<T_o$, the lattice parameters show   
only a small, linear $T$ dependence.    
The existence of ferromagnetism only for $\gamma > 162.8^\circ$    
is consistent with LCO thin film studies.~\cite{fapssl08}   
Furthermore, recent band-structure calculations~\cite{lh13} indicate that   
LCO is magnetic only for $\delta y < 0.52$, where $\delta y$ is a measure of %%%  
the rhombohedral distortion of the lattice.  For the measured %%%  
LCO lattice parameters, $\gamma_c$ corresponds to $\delta y = 0.53$, %%%  
which indicates excellent agreement between the calculations   
and experiments for bulk LCO.     
Thin film studies ~\cite{fdaeaeksgl09,rhsd10,pbsskzwld11,prkkmsfw09,fms08,srkkmsw12,fapssl08,hrsd09,fpssnssmrl07,spd12} have also found    
supporting evidence for a lattice distortion causing ferromagnetism.     
LCO material deposited on substrates that result in tensile strain %%% ambiguous wording 
show ferromagnetic order   
below $T \approx 87$~K, and the strength of the net ferromagnetic moment   
increases with the value of $\gamma$. %resulting from the strain   
Based on the results of experiments on bulk LCO and LCO films grown on   
various substrates,   
a model has been proposed in which surface-induced lattice   
stress increases $\gamma$ near the surfaces.~\cite{dbbycfb13} This %%%   
induces a transition to long-range ferromagnetic order at $T_C$   
throughout the LCO lattice.  For moments far from the surfaces,   
$\gamma$ becomes lower than $\gamma_c$ below $37$~K in bulk LCO. %changed 40 to 37  
At these low temperatures the lattice loses ferromagnetic order, %%%  
except near the surfaces where $\gamma > \gamma_c$. %%%  
   
We extend our previous study on bulk LCO to nanoparticles    
using the same neutron scattering and magnetometry techniques. The lattice    
parameters for both materials were determined over a range of    
temperatures $10 \le T \le 300$~K, and the field-cooled and zero-field-cooled   
magnetization was tracked in fields of $20$ Oe and $60$ kOe.     
We show that nanoparticles of LCO exhibit a phase transition    
very close to the transition temperature %$T_C \approx 87$~K (feel like we've said this enough) 
found in bulk LCO.  However, the magnetic order persists to low $T$,    
sharply contrasting the bulk behavior where the ferromagnetism   
collapses below $37$~K.  The ferromagnetic moment is much larger than    
in the bulk particles and continues to increase as the    
temperature decreases.   
We will show that these behaviors can be understood with the    
same Co-O-Co bond angle and surface-induced distortion model    
developed for the bulk behavior.       
   
LCO  nanoparticles were synthesized using the amorphous    
heteronuclear complex DTPA as a precursor,~\cite{ftyjyc00} using a method   
similar to that described in Ref.~\cite{jbsbamz09}. A NaOH    
solution at 1.0 M concentration was added by drops to a mixture solution of    
$\rm La(NO_3)_3$ and $\rm Co(NO_3)_3$ to prepare fresh hydroxides.    
A stoichiometric amount of NaOH was used to ensure complete reaction of the     
metal cations. The excess Na ions were then removed via    
dialysis over approximately 24 hours. Equimolar amounts of DTPA    
were then added to the metal hydroxides to synthesize the complex precursor.    
The mixture was stirred as it was heated to 80$^o$C. The resulting    
transparent solution was vaporized slowly at 80$^o$C until a dark purple   
resin-like gel formed. This precursor was decomposed in air at 350$^o$C     
for 1.5 hours to burn off the organic components. The resulting ash-like    
material was then heated at a calcination temperature of 620$^o$C for 4 hours.  

Zero-field neutron diffraction measurements were carried out with      
the US/Japan wide-angle neutron diffractometer (WAND) at the Oak Ridge    
National Laboratory High Flux Isotope Reactor with $\lambda$ = 1.48~\AA \      
using vanadium sample cans.      
Rietveld refinements were performed using FullProf.~\cite{r90}   
Small-angle x-ray scattering (SAXS) measurements to determine nanoparticle   
size and x-ray diffraction measurements were performed    
using a Rigaku SmartLab x-ray diffractometer.   
The magnetization, $M(T)$, was  measured for $H \le 60$~kOe using a      
Quantum Design SQUID magnetometer.    
       
\begin{figure}       
\includegraphics[height=3.4in, angle=0]{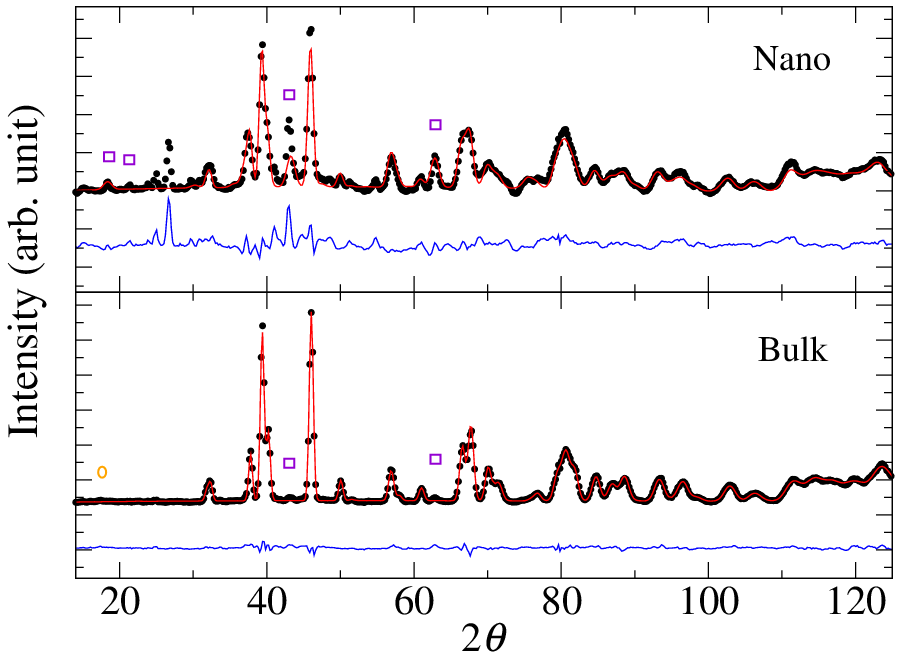}       
\caption       
{Neutron diffraction intensity vs $2\theta$ with FullProf refinements    
using the R$\overline{3}$c perovskite structure and      
differences for nanoparticles (upper) and LCO bulk (lower) at $T=10$~K.    
Nanoparticles show impurity peaks including those corresponding to      
$\rm Co_3O_4$ (box), which were fitted.      
A small amount of $\rm Co_3O_4$ was observed for the bulk and      
a small peak was observed at the antiferromagnetic position of     
$\rm CoO$ (circle).   
%these peaks were too small to be fit accurately,    
%so the CoO phases were not included in the bulk fits.  %%% don't really need this  
\label{fig:LCOnanobulk_refine}       
}       
\end{figure}      
       
The neutron diffraction intensity is shown versus $2\theta$ for    
the bulk and nanoparticle powders in Fig.~\ref{fig:LCOnanobulk_refine} at $T = 10$~K.       
Refinements were more difficult for the nanoparticles than for the bulk,       
likely due to a distribution of small, non-spherical particles which results   
in wider peaks that are difficult to model precisely.   
Noticeable Bragg peaks correspond to impurity phases   
that result from the low firing temperature used to   
create small LCO particles. The impurity phases will be addressed   
further below.   
SAXS measurements yield an average agglomerate particle size of       
75 nm with standard deviation of 21 nm.  A Debye-Scherrer analysis   
of the x-ray diffraction data yields LCO nanoparticle crystal sizes    
between 25-40 nm.   
%using the growth procedure described above.   
Hence, we conclude that the LCO    
nanoparticle crystals agglomerate into larger particles during the   
heating process.  No evidence of Bragg magnetic peaks in bulk or nanoparticles   
has been observed in experiments at WAND, which may be a consequence of the   
weak net FM moment.  

The diffraction patterns for both the bulk and nanoparticle LCO    
show small peaks corresponding to a $\rm Co_3O_4$ phase. The CoO %%%   
phase observed in the bulk was not evident in the nanoparticles. %%%  
Refinements indicate a weight     
fraction of $< 3.5\%$ $\rm Co_3O_4$ in the nanoparticles.    
The effect of these phases on the bulk has been previously    
discussed~\cite{dbbycfb13} and it applies similarly to    
the nanoparticles. In addition to the $\rm Co_3O_4$ phase, there are %%%   
also some Bragg peaks that were not identified as %%%  
expected oxides of La or Co. When comparing the bulk to the nanoparticle 
patterns, the lattice structures of these phases    
appear distinct from that of the LCO bulk. Unless there were a significant 
amount of impurity peaks directly overlapping the LCO peaks, we would 
not expect them to significantly %%%   
affect the quality of the LCO parameter refinements. The  
magnetization data do not show any unusual behavior which    
can be readily attributed to these impurity phases, although calculations  
of the average effective Co moment may reflect the presence of non-LCO phases.      
   
\begin{figure}      
\includegraphics[height=3.4in, angle=0]{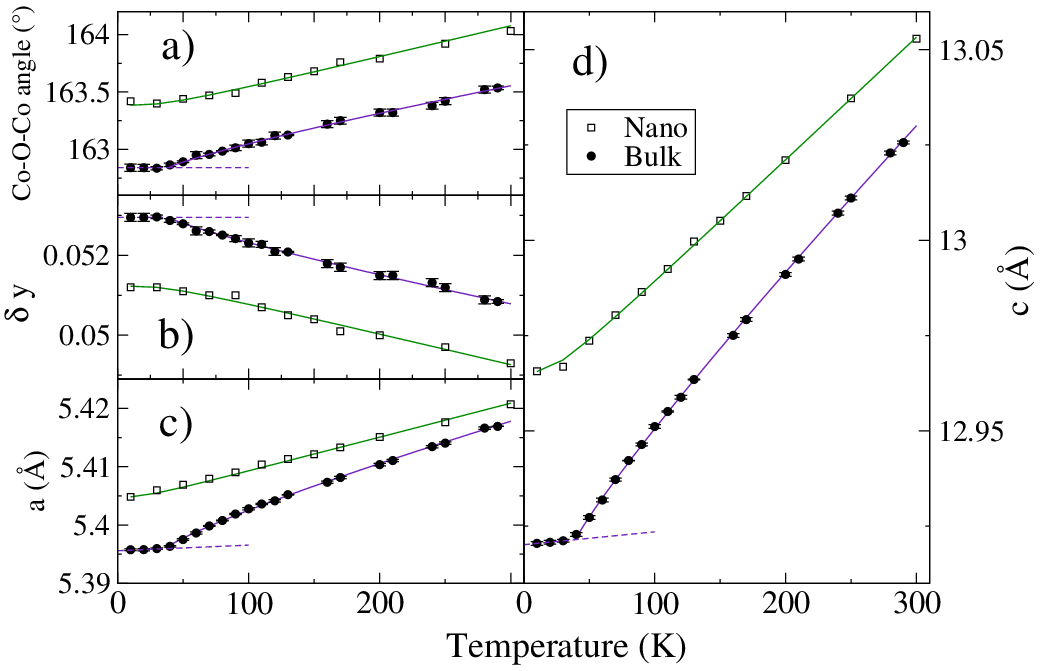}      
\caption      
{Lattice parameters $a$, $c$, $\delta y$, and the Co-O-Co angle ($\gamma$) for LCO      
nanoparticles and bulk powder.  Solid curves for the bulk represent power-law    
behavior in ($T-T_o$) and dashed lines linear behavior. The solid curves  %%% removed the cite  
for the nanoparticles are fits using the Gruneisen expression with the Einstein model   
power-law behavior is not apparent and $\gamma > \gamma_c$ for all $T$ for the nanoparticle case.    
\label{fig:LCObulknano_params}      
}      
\end{figure}

Fitted lattice parameters $a$ and $c$ for the hexagonal   
unit cell are shown for $0<T<300$~K in  Fig. \ref{fig:LCObulknano_params}   
and are compared to previous results obtained for bulk LCO.      
The Co-O-Co bond angle ($\gamma$), and the parameter      
$\delta y = \frac{d}{a}\cos(\gamma/2)$, where $d$ is the Co-O bond length,      
are also shown. The parameter $\delta y$ describes the deviation of   
the oxygen position from the straight line connecting neighboring Co   
ions and characterizes the amount of rhombohedral distortion   
of the lattice.~\cite{rkfk99,dbbycfb13,mkfoiyk99} Although $\delta y$ is a 
derived quantity from $a$ and $\gamma$, we include it in the fits so as to be 
consistent with other relevant papers which interpret their results in 
the context of $\delta y$.~\cite{lh13, dbbycfb13}    
   
Contrasting the sharp changes observed near $T_o \approx 37$~K in bulk LCO,  %%%37 instead of 39  
$\delta y$, $\gamma$, $a$ and $c$ show no abrupt change in      
slope in the nanoparticles. The value of $\gamma$ observed for the nanoparticles    
remains well above $\gamma_c$ throughout the entire range of temperatures studied.      
   
\begin{table}      
\caption{The fits to the nanoparticle lattice parameters    
using Eq.~\ref{eq:Grun}. The errors in $\alpha$ and $T_E$   
are 1$\%$ of the given values.   
\label{Table:nanoparams}      
}      
\begin{tabular}{l*{5}{c}r}      
  Parameter         & $y(0) $ && $\alpha$ && $T_E (\rm K) $ \\      
  \hline      
  a                 &  5.405(4) && 2.7$\ee{-4}$ && 50.0 \\ %%%errors - estimated because xmgrace doesn't provide them   
  c                 &  12.966(1) &&  7.3$\ee{-4}$ && 58.6  \\      
  $\delta y$        &  0.0512(1)  &&  -7.5$\ee{-3}$ && 98.7 \\      
  Co-O-Co ($\gamma$)           &  163.384(4)  &&  8.0$\ee{-4}$ && 96.1 \\      
\end{tabular}      
\end{table}     
   
The temperature dependence of the nanoparticle lattice parameters were fit using   
the Gruneisen expression with the Einstein    
model for thermal lattice expansion,   
   
\begin{equation}      
y(T) = y(0)[1 + \alpha(\coth(\frac{T_E}{2T}) - 1)] \quad ,      
\label{eq:Grun}      
\end{equation}      
      
\noindent where the lattice parameter being fit is $y(T)$, $y(0)$ is its    
value at $T=0$, $T_E$ is the Einstein temperature, and $\alpha$ is the    
thermal expansion coefficient for $T >> T_E$.~\cite{crbmrsrfg99}    
The lattice parameters are well fit by the thermal expansion model,   
but $T_E$ is rather small and inconsistent among the various   
parameters.  It is not surprising that data   
for lattice parameters that exhibit no sharp features   
can be fit by the simple temperature dependence of Eq.~\ref{eq:Grun}.   
No power-law behavior or phase transition is observed for   
nanoparticles, unlike the bulk LCO.      
      
\begin{figure}      
\includegraphics[height=3.4in, angle=0]{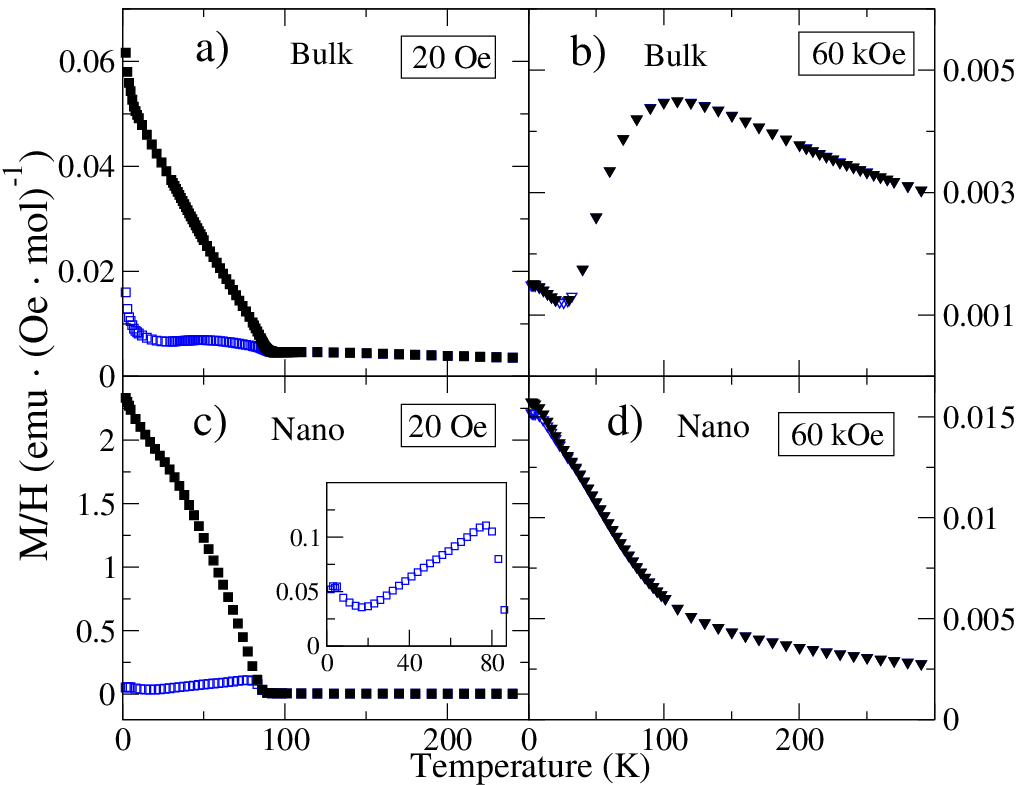}      
\caption      
{$M/H$ vs.\ $T$ for nanoparticle and bulk LCO.      
The magnetic behavior is shown in a) and c) for $H=20 \pm 1$~Oe and b)    
and d) for 60~kOe. The ZFC data are shown by open symbols and the FC data    
by closed symbols. ZFC and FC data overlap closely for the 60 kOe measurements.   
The inset expands the nanoparticle ZFC data. Note the    
different vertical scales for the bulk and nanoparticles, and in c) the 0 is 
offset for clarity. The behaviors    
for $T>100$~K are similar in magnitude and field independent for    
all cases.   
\label{fig:LCO_bulknanomag}      
}      
\end{figure}

$M/H$ vs $T$ for the $\rm LaCoO_3$ bulk and nanoparticles at $H=20 \pm 1$~Oe is shown in % discussing 20 Oe   
Fig.~\ref{fig:LCO_bulknanomag} a) and c). The nanoparticle magnetic phase transition % discussing FC   
occurs at nearly the same temperature as the bulk, indicating that the  %%%  
magnetism in both materials is of the same origin. %%%  
The field-cooled (FC) and zero-field-cooled (ZFC) behaviors    
for both materials diverge just below $T_C$.  The nanoparticles show a much larger    
FC moment and a much larger difference between the FC and ZFC behaviors. The magnetization   
below a FM phase transition   
is expected to exhibit the power law behavior $t^\alpha$ as the individual 
moments align with each other, where $t=(T_C-T)/T_C$ ($t << 1$) and 
$\alpha < 1/2$.  Hence, we normally expect significant curvature   
below $T_C$. However, the net moment in bulk LCO
shows little curvature ($\alpha \approx 1$). This likely reflects the very small
net FM moment as well as either a decrease in
the local moment size or the strength of the interaction between moments
as $\gamma$ decreases with temperature towards $T_o$.
Although the nanoparticle magnetization exhibits greater curvature with a
value $\alpha \approx 0.8$ for fits
over the range $32 < T < 84$~K, the curvature is still not as large as normally
expected for a FM phase transition.  The nanoparticle
moments and interactions vary less than in the bulk, but
the net FM moment remains small (though much larger than in the bulk). 
Overall, the rounding of the transition near $T_C$ and significant non-critical
contributions well below $T_C$ preclude meaningful fits to the
critical behavior.  

%However, the bulk behavior has little curvature ($\alpha \approx 1$), which is   
%likely the influence of the decrease of the ferromagnetic moment with $\gamma$ as the %%% 
%transition at $T_o$ is approached.   
%The nanoparticle magnetization exhibits greater curvature with a fitted value $\alpha = 0.8$ %%%  
%over $32 < T < 84$~K, which is likely a result of the much larger ferromagnetic 
%moment and relatively smaller noncritical contributions.
   
%%% see the figure ``LCOnano_magfig_fit.ps''   
%Add more, cite other papers, flesh out where other PM magnetization might come in    
   
The shapes of the ZFC curves in the bulk and nanoparticles show   
a similar paramagnetic-like tail below $T=20$~K, except that   
the nanoparticle moment appears to saturate at $0.054$ $\rm \frac{emu}{Oe \cdot mol}$.   
Although the individual nanoparticles are fairly well ordered, there are likely moments   
at grain boundaries that contribute to the paramagnetic tail.   
The nanoparticle ZFC magnetization between $T=20$~K and $T_C$ is   
much larger than that of the bulk and probably represents   
randomly oriented net ferromagnetic moments   
of the nanoparticle grains which are not readily aligned by the applied field.   
%I think the descriptions before were too detailed, speculative, and maybe not important   
%enough to spend too much time on.  Let's discuss the ideas more.   
%Although this all makes sense to me, I don't like how these mechanisms are different for the bulk    
%and the nano...should rewrite this eventually and make it more cohesive. Cite more stuff   

\begin{figure}       
\includegraphics[width=2.9in, angle=0]{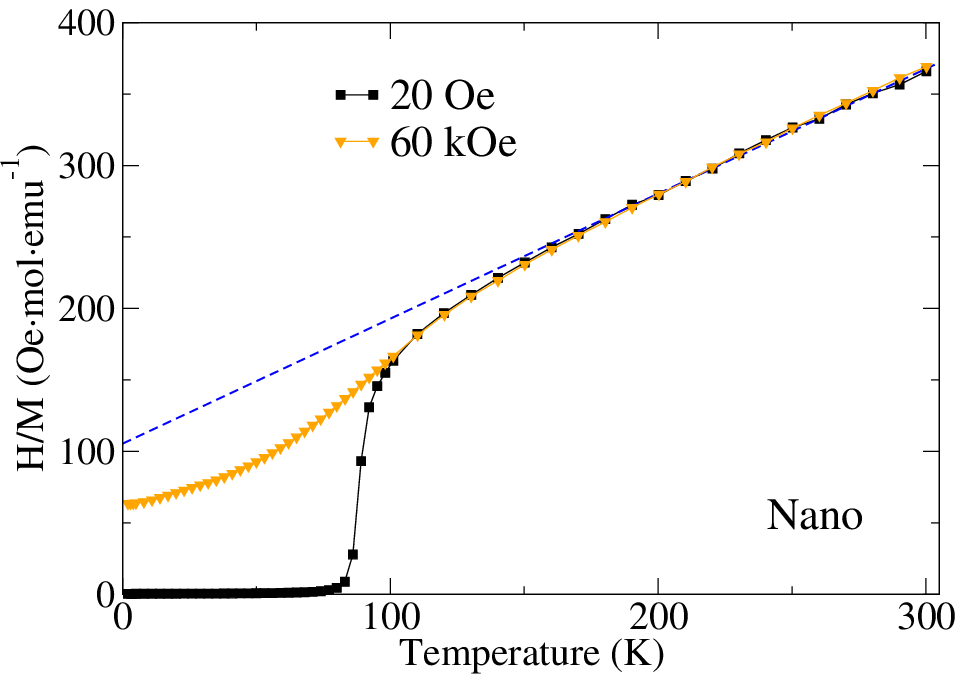}       
\caption       
{$H/M$ vs. $T$ for nanoparticle LCO at $H=20$~Oe and $60$~kOe.    
The linear region in both fields is fit to Curie-Weiss behavior (dashed line).       
\label{fig:LCObulk_invMH}       
}       
\end{figure}

The $M/H$ nanoparticle behavior for $H=60$~kOe sharply contrasts that of the bulk particles.      
Whereas the bulk magnetization decreases with $T$ below $90$~K, with a minimum just below    
$40$~K (Fig.~\ref{fig:LCO_bulknanomag} b), the nanoparticle moment increases    
monotonically as $T$ dereases over the entire temperature range (Fig.~\ref{fig:LCO_bulknanomag} d).   
%Actually, I would say that the transition is still at H=0, so it is not   
%that surprising that the transition looks rounded in large fields.   
%The external magnetic field overwhelms the ferromagnetic domains in    
%the nanoparticles, so the behavior appears paramagnetic.   
The transition is rounded for $H=60$~kOe because the critical   
point is at $H=0$. However, there is a slight inflection point   
near $T_C$, which is more clear from $H/M$ versus $T$ data shown in Fig.~\ref{fig:LCObulk_invMH}.   
This indicates a strong influence of the ferromagnetic interactions in high field.   
The bulk data indicate no such signifcant FM interaction at high fields, consistent with   
the small net ferromagnetic moment observed in the bulk and the disappearance of the   
ferromagnetism below the transition at $T_o$.   
  
Figure~\ref{fig:LCObulk_invMH} shows $H/M$ versus $T$       
for the nanoparticles at $H=60$~kOe and $20$~Oe.     
The steep drop in $H/M$ with decreasing temperature  
beginning at $T = 100$~K is indicative of ferromagnetism %cite something for this!   
for $20$~Oe.  The slight dip at the same temperature for $60$~kOe    
demonstrates that the ferromagnetic interactions are still significant.   
Data at both fields show nearly the same %significant %%% why the word significant?   
paramagnetic behavior above $170$~K.   
The straight line fit for $170 \leq T \leq 300$~K is     
interpreted as Curie-Weiss (CW) behavior      
using $H/M=1/\chi = (T-\theta_{CW})/C$. We obtain a CW temperature  
$\theta_{CW}=-121(3)$~K and Curie constant $C=1.14(1)$~emu$\cdot$K$\cdot$mol$^{-1}$ for the    
nanoparticles. This value of $\theta_{CW}$ is consistent with     
$T_C=87$~K (CW calculations yield upper limits for transition temperatures)    
and suggests dominant antiferromagnetic interactions above $170$~K.   
Both values are similar to those obtained for bulk LCO, where   
$\theta_{CW} = -138(4)$ and $C = 1.31(2)$.   
Above the magnetic transition at $87$~K, the effective moment is   
$\mu_{eff} = 3.03(2)$ $\mu_{B}$ per Co ion. This is  
smaller than the bulk value of $3.23(2)$ $\mu_B$. The less negative %%% 
$\theta_{CW}$ and smaller $\mu_{eff}$ perhaps indicate 
that fewer moments are participating due to the less perfect  
crystallinity of the nanoparticle powder compared to the bulk.  
The impurities present in the nanoparticles may also contribute to  
an overestimation of the number of participating Co ions. %what values of mu_eff did others find? 
A larger FM interaction in the nanoparticles could also result  
in a less negative $\theta_{CW}$. Notably, Fita \textit{et al.} found  
an even smaller value of $\mu_{eff} = 2.44 \mu_B$ for their  
LCO nanoparticles, as well as a less negative $\theta_{CW}$ of  
-48 K.~\cite{fmmpwtvhvg08}  
   
%It is notable that the paramagnetic behavior indicates antiferromagnetic    
%interactions, but that below $T_C$ the nanoparticles exhibit ferromagnetic    
%behavior. % Why is that? Frustration prevents antiferromagnetism and so ferromagnetic long-range order   
%becomes possible.   
   
In several previous studies, the ferromagnetism in LCO nanoparticles has been   
attributed to ferromagnetic ordering of the surface,~\cite{yzg04_a,httski07}   
surface-induced lattice strain,~\cite{fmmpwtvhvg08} and unit-cell expansion.~\cite{wzwf12,zhzgzs09}   
%categorized according to what they stated as the main cause of the FM in the paper  
Yan \textit{et al.}~\cite{yzg04_a} found that the magnetic susceptibility of their   
samples increased as the surface-to-volume ratio increased and attributed   
this to localized spins on the surface of the material. Although they ruled   
out double exchange between Co(III) and Co(IV) atoms as the mechanism for   
the surface ferromagnetism, their study was inconclusive as to the origin   
of the ferromagnetic interaction. Harada \textit{et al.}~\cite{httski07} found similar   
results; the magnetization increased with decreasing particle   
size and they suggested the source to be chemisorption of oxygen atoms at the surface.   
Again, the mechanism leading to ferromagnetism was not made explicit.   
Fita \textit{et al.}~\cite{fmmpwtvhvg08} examined the lattice parameters of LCO   
nanoparticles and found that they increase with decreasing particle size   
%(consistent with our observations comparing bulk and nanoparticle LCO ~\cite{dbbycfb13}).    
but did not identify surfaces as the source of ferromagnetism.  Instead, they   
pointed to the surface-induced lattice expansion which persists throughout   
the material as the cause.  In the above studies, LCO crystals and powders were   
synthesized using several different methods: floating-zone single-crystal   
synthesis, solid-state reaction, crushing the single-crystal into a powder,   
and a citric acid solution method. It is interesting to note that in all but one   
case, the magnetic susceptibility increased monotonically with surface   
area and the transition temperature remained fairly constant at $T_C \approx 85$~K.   
Wei \textit{et al.}~\cite{wzwf12} found that their nanoparticles showed a decrease   
in magnetization and $T_C$ with particle size, despite also noting a similar increase   
in lattice parameters with the smaller sizes. It is possible that the citric acid sol-gel   
method employed in this case differed slightly from the one used by   
Harada \textit{et al.},~\cite{httski07} but it is unclear how that would produce   
results that differ qualitatively from the others.    
   
Tensile stress from substrates allows LCO thin films to order %%%  
ferromagnetically. The resulting FM order is found throughout the material %%%  
for films of order 100 nm~\cite{fdaeaeksgl09} and the net moment increases with %%%   
film thickness.~\cite{fpssnssmrl07}  In nanoparticles, %%%   
all of the LCO material is well within 100 nm of a surface. %%%  
Hence, tensile stress from these surfaces should   
result in a larger average $\gamma$ for the particles, although 
the exact process by which the strain changes $\gamma$ requires further investigation. 
Our experiments show that $\gamma$ does remain well above $\gamma_c$   
for all $T$ (Fig.~\ref{fig:LCObulknano_params}), which is consistent   
with the net FM moment being much larger than in bulk LCO and   
not collapsing below $T=37$~K.  Note that the lowest value of $\gamma$ %changed 40 to 37, ``average'' -> ``lowest''  
in the nanoparticles is comparable to the bulk value for $T \approx 200$~K,   
so we would expect the system to remain ferromagnetic for $T<87$~K in   
small $H$. 
This model, which correlates the stability of ferromagnetism with the   
stress-induced increase in $\gamma$, is consistent with the observations   
made by most of studies noted above.  
The ferromagnetic transition at $T=87$~K is a result of the   
stress at surfaces, but long-range ferromagnetic order takes place throughout the LCO lattice.   
Only for bulk LCO is the long-range order observed to disappear below $T=40$~K because only   
in this system is $\gamma$ observed to decrease below $\gamma_c$.  
 
Although $\gamma$ is a  
useful parameter by which to gauge the degree of magnetism in a sample, the mechanism by which  
it controls the ferromagnetism remains unclear. Calculations by Lee and Harmon indicate  
$\delta y$ (closely related to $\gamma$) to be the controlling parameter in LCO and  
note that the amount of rhombohedral lattice distortion determines whether the  
ground state is magnetic.\cite{lh13} According to this model, the degree of  
orbital overlap between the Co and O ions affects the balance between  
the repulsive Coulomb interaction and the exchange interaction: more distortion  
(and less overlap) leads to a non-magnetic ground state.  
Goodenough proposes that the  
local configurations of low-spin and high-spin Co$^{3+}$ and Co$^{4+}$ ions  
result in areas of ferromagnetism, antiferromagnetism, and paramagnetism.\cite{g58}  
It is also conceivable in this interpretation that the Co-O-Co bond angle affects the  
spin state of the Co ions or groups of ions by allowing for increased  
or decreased overlap of the cobalt and oxyen orbitals. %%% FINISH THIS 
  
%DISCUSS THEORY WITH REGARDS TO IMPURITIES AND SURFACES HERE (Co3O4, other things, surfaces)   
%Mention tensile vs. compressive strain (films)   
%Mention how the other nano papers relate to us   
%Discuss Co-O-Co bond angle here   
    
%Paragraph below should actually be a summary, not introduce new ideas   
In summary, we have shown that,   
for $T>87$~K, the interaction strengths and paramagnetic behavior %%% why 100 K, not 87?  
are comparable in bulk and nanoparticle LCO.  On the other hand, the observed  
magnetic behaviors observed for $T<87$~K in LCO bulk and nanoparticles %%%  
are very different.  Nevertheless, the magnetic behaviors of bulk and nanoparticles  
can be modeled within the same framework; the interactions between spins in LCO are strongly    
dependent on the Co-O-Co bond angle $\gamma$.  Magnetic    
order can only be sustained when $\gamma$ is larger than a critical    
value $\gamma_c =162.8$~$^\circ$.  In bulk LCO, this occurs only for    
$T>T_o$, where $T_o \approx 37$~K.  For $T<T_o$, magnetism in    
bulk LCO is associated only with regions of tensile stress near   
surfaces and interfaces with impurity phases.  In    
LCO thin films and nanoparticles, all moments are near to surfaces   
so $\gamma$ is always larger than $\gamma_c$ and long-range ferromagnetic order   
is present for all $T<87$~K.   
      
We thank Y. Abdollahian, F. Bridges, C. de la Cruz,     
A. Elvin, B. Harmon, J. Howe, S. Shastry, and N. Sundaram for     
helpful discussions and$/$or assistance with measurements.      
The work at ORNL is supported by the DOE BES Office    
of Scientific User Facilities.    
Work at Lawrence Berkeley National Laboratory was      
supported by the Director, Office of Science (OS),     
Office of Basic Energy Sciences (OBES), of the U.S. Department of      
Energy (DOE) under Contract No. DE-AC02-05CH11231.     
Some X-ray data in this work were recorded on an instrument   
supported by the NSF Major Research Instrumentation (MRI)   
Program under Grant DMR-1126845.   

\bibliography{magnetism.bib}       
       
\end{document}